\documentclass[eqsecnum,floats,aps,nofootinbib,showpacs]{revtex4}
\usepackage{latexsym}
\usepackage{amssymb}

\renewcommand{\d}{\mathrm{d}}
\renewcommand{\l}{\left(}
\renewcommand{\r}{\right)}
\usepackage{amsmath, amsthm, amssymb}

\def\be{\begin{equation}}
\def\ee{\end{equation}}
\def\beq{\begin{equation*}}
\def\eeq{\end{equation*}}
\def\ba{\begin{aligned}}
\def\ea{\end{aligned}}

\def\ov{\overline}
\def\w{\wedge}

\begin{document}
\title{Nontrivial realization of the space--time translations in the theory of quantum fields}

\author {Marcin Ka\'zmierczak}
\email{marcin.kazmierczak@fuw.edu.pl}
\affiliation{Institute of Theoretical Physics, Uniwersytet Warszawski, Ho\.{z}a 69, 00-681 Warszawa, Poland} 

\begin{abstract}
In standard quantum field theory, the one--particle states are classified by unitary representations of the Poincar{\'e} group, whereas the causal fields' classification employs the finite dimensional (non--unitary) representations of the (homogeneous) Lorentz group. 
A natural question arises -- why the fields are not allowed to transform nontrivially under translations? 
We investigate this issue by considering the fields that transform under the full representation of the Poincar{\'e} group. It follows that such fields can be consistently constructed, although the Lagrangians that describe them necessarily exhibit explicit dependence on the space--time coordinates. The two examples of the Poincar\'e--spinor and the Poincar{\'e}--vector fields are considered in details. The inclusion of Yang--Mills type interactions is considered on the simplest example of the $U(1)$ gauge theory. The generalization to the non--abelian case is straightforward so long as the action of the gauge group on fields is independent of the action of the Poincar\'e group. This is the case for all the known interactions but gravity.
\end{abstract}
\maketitle

\section{Introduction}\label{intro}

In the conventional approach to quantum field theory (QFT), the Hilbert spaces of one--particle states are constructed by considering irreducible unitary representations of the Poincar{\'e} group
\footnote{
More precisely, its universal covering group, if spinor representations are to be included. This remark should be understood to hold throughout the paper.
}
derived by induction from irreducible representations of the little group which are labeled by spin (massive case) or helicity (massless case). Then the quantum causal fields are introduced that are classified by finite--dimensional representations of the (homogeneous) Lorentz group. Any such representation, acting on the space of fields,  needs to be connected to some unitary representation of the Poincar{\'e} group, acting on the Fock space of many--particle states, via Weinberg consistency conditions. For a fixed representation of the Lorentz group, the consistency conditions restrict the set of allowable representations of the Poincar{\'e} group on the Fock space. 
Choosing one of these representations and solving the conditions leads to the form of the field amplitudes that correspond to a unique  value of spin and can be interpreted as describing a particular kind of particles. Further restrictions on the amplitudes follow from causality conditions, charge conservation, self--adjointness of the interaction density and possibly discrete symmetries. The fields thus constructed obey Lorentz invariant equations which can be given Lagrangian formulation. Then, the conserved currents of physical interest can be inspected by means of Noether theorem, the interaction terms and the $S$--matrix can be constructed. See the classical reference \cite{Wein} for a review of this approach to QFT.

In fact, the Lagrangians of such theories are invariant under the global action of the Poincar{\'e} group where the translations act trivially. The Lorentz group, on the other hand, can act trivially (e.g. for a scalar field) or non--trivially (e.g. for a vector or a spinor field). 
 A natural question is in order:  what is the reason for classifying causal fields by representations of the Lorentz group only and not the full Poincar{\'e} group? In other words, why are all the finite--dimensional representations of the Poincar{\'e} group that classify the fields non--faithful? 

This question if of particular interest in the connection with the attempts to describe all the ineractions within a conceptually unified framework, at least on the classical level. Assume that gravity is sufficiently small to be neglected in some sort of experiments. Then special relativity implies that every theory accounting for the results of these experiments need to be invariant under the global action of the Poincar{\'e} group corresponding to the passage from one inertial observer to another. The basic idea of the Poincar{\'e} gauge theory of gravity (PGT) is to introduce gravity by localization of this fundamental symmetry. Since the pioneering work of Yang and Mills on the strong interactions \cite{YaMi}, such an approach to the description of interactions that is based on localization of global symmetries has proved extremely successful in the non--gravitational sector. The idea of describing gravity in a similar way has now a long history. Initially, gravity was viewed as a gauge theory of the Lorentz group by Utiyama \cite{Ut}. Then Kibble \cite{Kib1} observed that promoting the whole Poincar{\'e} group to the gauge group has a lot of  advantages. Among them, one  is not forced to introduce a cotetrad on space--time {\it ad hoc} -- it can be related to the translational gauge fields and derived via the localization procedure, in much the same way as a space--time connection. Further investigations of the idea where made (see e.g. \cite{HHKN} for a review), but the translational and homogeneous parts of the group were not really treated on the same footing until the work of Grignani and Nardelli \cite{GN}, where the authors realized that 
only after additional fields are introduced on space--time can the theory be cast into the form that is truly similar to the geometric setting of standard Yang--Mills theories. These fields, called the {\it Poincar{\'e} coordinates}, transform as Poincar{\'e} vectors under gauge transformations. In fact, their geometric interpretation in terms of the theory of connections on a principal fiber bundle of affine frames was given much earlier by Trautman \cite{T1}, but the physical interpretation in the usual formulation of the theory in terms of sections of the bundle was not discussed there. What is more, the complete affine group was considered as a gauge group, rather than merely its Poincar{\'e} subgroup. For an exhaustive review of possible approaches to the gauge formulation of gravity, see \cite{HCMN}. It is important to remember that the Poincar{\'e} group remains the most natural candidate for the gauge group from the physical point of view because of its relation to the principle of equivalence of special relativity. 

However, this approach to the theory of gravity can be criticized. One of the points that are rised by the oponents of PGT is that one does not normally see the translations acting. The translational gauge fields are hidden within the cotetrad that transforms homogeneously under the action of the Poincar\'e group. What is more, all the physically important matter fields are translational scalars. Therefore, their gravitational covariant derivatives do contain generically the Lorentz gauge fields that give rise to the space--time connection, but do not contain the translational gauge fields. These observations may lead to the conclusion that gauging translations is a somewhat artifficial procedure and one should rather aim to formulate gravity as a gauge theory of the Lorentz group, following Utiyama, or abandon the gauge approach at all. However, these doubts concerning the relevance of PGT would dispel if the quantum fields that transform under faithful representations of the Poincar\'e group were discovered in nature. But is it possible to construct a quantum theory of such fields consistently? If it is, will they differ physically from standard quantum fields? These questions are of particular importance at the time of LHC beginning to collect new data. If the physical properties of Poincar\'e quantum fields are not worked out, it may well be the case that thay will be discovered but not recognized. In order to avoid misinterpretation of the data, it is important to extend quantum field theory to include fields that transform nontrivially under translations. 
Although the classical theory of such fields attracted a lot of interest in the course of time \cite{GN}\cite{Lec1}\cite{TT1}\cite{TT2}, with the particular emphasis laid on Poincar\'e spinors, the quantum theory of such fields was not satisfactorily investigated, according to the best knowledge of the author of this article.

We will discuss such a theory in Section \ref{free}. 
It will appear that the quantum approach to the problem gives a better understanding and deeper insight into some aspects of the corresponding classical theory that were not satisfactorily addressed in the hitherto accounts on Poincar\'e fields. In section \ref{Int}, the construction of interaction terms will be addressed and the simplest gauge theory of electromagnetism will be considered. The generalisation to the case of non--abelian gauge theories whose gauge groups act on fields independently on the action of the Poincar\'e group is straightforward and the case of gravity is going to be considered in the forthcoming paper. In section \ref{conc} we draw the conclusions.

\section{Free causal fields in quantum field theory with nontrivial translations}\label{free}

\subsection{General formalism}\label{general}

In this section we shall parallel the basic derivations of \cite{Wein} allowing for the presence of nontrivial generators of momenta in the representations that classify the 
quantum fields.
Let $U(\Lambda,b)=U(b)U(\Lambda)$, $U(b)=\exp\l{ib\cdot P}\r, \ U\l{\Lambda(\varepsilon)}\r=\exp\l{\frac{i}{2}\varepsilon_{ab}J^{ab}}\r$ be the irreducible unitary representation of the universal covering of the Poincar{\'e} group. Here $P^a, J^{ab}$ are the self--adjoint generators of translations and Lorentz rotations belonging to the corresponding representation of the Poincar{\'e} algebra, $\varepsilon\equiv\l{{\varepsilon^a}_b}\r\in so(1,3)$ is such that $\Lambda\l{\varepsilon}\r{^a}_b={\delta^a}_b+{\varepsilon^a}_b+\dots$, $b\equiv\l{b^a}\r$ are the parameters of space--time translation and $\cdot$ denotes the Minkowski product, $b\cdot P=\eta_{ab}b^aP^b$, $\eta=diag(1,-1,-1,-1)$. We shall restrict ourselves to the massive representations for which $P\cdot P=m^2$, where $m>0$ is interpreted as a mass of a particle. Then the little group that leaves the standard momentum $k=(m,0,0,0)$ invariant is $SO(3)$ and its universal covering is $SU(2)$. Let $D^j$ denote the irreducible unitary representations of $SU(2)$ labeled by spin $j=0,\frac{1}{2},1\dots$\footnote{
Note that the expressions of the form $D^j(R)$ for $R\in SO(3)$, which we shall use for simplicity, are multi--valued for half--integer values of $j$.
}. Let $\Psi_{p,\sigma}$ be the (distributional) basis of the space of one--particle states with well established value of momentum $p$ and the projection of spin on the third spatial axis in the rest frame $\sigma$. The action of the representation $U$ on this basis is
\be\label{Utr}
U(\Lambda,b)\Psi_{p,\sigma}=e^{ib\cdot\Lambda
  p}D^j\l{W_{\Lambda,p}}\r_{\sigma'\sigma}
\Psi_{\Lambda p,\sigma'} \quad ,
\ee
where
$L_p\in SO(1,3)$ is the standard bust, $L_pk=p$, given explicitly by 
\be\label{bust}
L_p=
\l
\begin{array}{cccc}
\frac{p^0}{m}\qquad & \frac{\mathbf{p}^T}{m} \\
\frac{\mathbf{p}}{m}\qquad & {\mathbf 1}_3+\frac{\mathbf{p}\mathbf{p}^T}{m(p^0+m)} 
\end{array}
\r
\ee
(think of $\mathbf{p}\in\mathbb{R}^3$ as a column matrix)
and 
$W_{\Lambda,p}={L_{\Lambda p}}^{-1}\Lambda L_p$ belongs to the little group of $k$. We shall follow the normalization convention according to which
$\Psi_{p,\sigma}=U(L_p)\Psi_{k,\sigma}$.
The one--particle states are created from vacuum according to $a^{\dag}_{p,\sigma}\Psi_0=\Psi_{p,\sigma}$,
where $\Psi_0$ represents the vacuum state, normalized to unity. The commutation relations for creation and annihilation operators are adopted in the form 
\be
[a_{p,\sigma},a^{\dag}{}_{p',\sigma'}]_{\mp}=(2\pi)^32p^0\delta({\bf
  p}-{\bf p}')\delta_{\sigma,\sigma'}\quad,
\quad [a_{p,\sigma},a_{p',\sigma'}]_{\mp}=[a^{\dag}_{p,\sigma},a^{\dag}_{p',\sigma'}]_{\mp}=0\quad,
\ee
where the upper sign in $\mp$ denotes the commutator and refers to bosons, while the lower one denotes the
anti--commutator and refers to fermions (in all the formulas the on--mass--shall condition $p^0=\sqrt{{\bf p}^2+m^2}$ should be understood to hold). Assuming the invariance of the vacuum $U(\Lambda,b)\Psi_0=\Psi_0$, one can derive from (\ref{Utr}) the transformation law for the creation operators
\be\label{atr}
U(\Lambda,b)a^{\dag}_{p,\sigma}U^{-1}(\Lambda,b)=e^{ib\cdot\Lambda
p}D^j\l{W_{\Lambda,p}}\r_{\sigma'\sigma}a^{\dag}_{\Lambda p,\sigma'}\quad,
\ee
from which the transformation law for the annihilation operators follows by conjugation.

Let us introduce creation and annihilation fields 
\be
\psi^{+}_l(x)=\int \, u_{l\sigma}(x,p)a_{p,\sigma} \, d\Gamma_p\quad , \quad
\psi^{-}_l(x)=\int \, v_{l\sigma}(x,p)a^{\dag}{}_{p,\sigma}\, d\Gamma_p \quad ,
\ee
where $d\Gamma_p=\frac{d^3p}{(2\pi)^32p^0}$ is the Lorentz--invariant measure.
We shall require that these fields satisfy the transformation law
\be\label{psitr}
U(\Lambda,b)\psi^{\pm}_l(x)U^{-1}(\Lambda,b)=\rho^{-1}_{ll'}(\Lambda,b)\psi^{\pm}_{l'}(\Lambda
x+b) \quad ,
\ee
where 
\be\label{Poinrep}
\ba
&\rho(\Lambda,b)=\rho(b)\rho(\Lambda)\quad,\quad \rho(b):=\rho\l{{\bf 1},b}\r=\exp\l{ib\cdot\mathbb{P}}\r\quad, \\
&\rho(\Lambda(\varepsilon)):=\rho\l{\Lambda(\varepsilon),0}\r=\exp\l{\frac{i}{2}\varepsilon_{ab}\mathbb{J}^{ab}}\r
\ea
\ee
is a finite--dimensional (non--unitary) representation of the Poincar{\'e} group. Thus the formula differs from the one considered in \cite{Wein} by the presence of the non--zero momentum generators $\mathbb{P}^{a}$. Note that for $\Lambda={\bf 1}$ and infinitesimal $b$ (\ref{psitr}) implies that
\be
[P_a,\psi^{\pm}(x)]_{-}=\l{-i\partial_a-\mathbb{P}_a}\r\psi^{\pm}(x)\quad .
\ee

Using (\ref{atr}), the consistency conditions relating the representations $U$ and $\rho$ can be derived
\be\label{war}
\ba
\rho^{-1}(\Lambda,b) u(\Lambda x+b,\Lambda p)&=e^{-ib\cdot\Lambda p} u(x,p){D^j}^{-1}\l{W_{\Lambda,p}}\r \quad , \\ 
\rho^{-1}(\Lambda,b) v(\Lambda x+b,\Lambda p)&=e^{ib\cdot\Lambda p} v(x,p){D^j}^T\l{W_{\Lambda,p}}\r \quad ,
\ea
\ee
where $u$ and $v$ denote matrices whose entries are $u_{l\sigma}$ and $v_{l\sigma}$ respectively and $^T$ stands for the transposition of a matrix. 
For pure translations one gets 
\be
u(x+b,p)=e^{-ib\cdot p}\rho(b)u(x,p) \quad , \quad v(x+b,p)=e^{ib\cdot p}\rho(b)v(x,p) \quad .
\ee
The solution is provided by the following form of the amplitudes $u$ and $v$
\be\label{uxvx}
u(x,p)=e^{-ip\cdot x}\rho(x)u(p) \quad, \quad v(x,p)=e^{ip\cdot
  x}\rho(x)v(p) \quad,
\ee
where $u(p)\equiv u(0,p)$ and $v(p)\equiv v(0,p)$ do not depend on $x$. 
Thus the fields $\psi^{\pm}(x)$ are not just the Fourier transforms, as in the conventional QFT.  Inserting (\ref{uxvx}) into (\ref{war}) and
employing the composition law
$\rho(\Lambda,a)\rho(\Lambda',a')=\rho(\Lambda\Lambda',\Lambda a'+a)$ we get the standard Weinberg conditions
\be
u(\Lambda p)=\rho(\Lambda)u(p){D^j}^{-1}\l{W_{\Lambda,p}}\r \quad, \quad
v(\Lambda p)=\rho(\Lambda)v(p){D^j}^{T}\l{W_{\Lambda,p}}\r \quad.
\ee
Hence, the $x$--independent parts of the amplitudes satisfy the conditions of standard theory. In particular, it follows that
\be\label{pchobr}
\ba
&u(p)=\rho(L_p)u(k) \quad, \quad v(p)=\rho(L_p)v(k) \quad, \\
&u(k)=\rho(R)u(k){D^j}^{-1}(R) \quad, \quad v(k)=\rho(R)v(k){D^j}^T(R) \quad,
\ea
\ee
for the standard momentum $k$ and any rotation $R$. Only the dependence of $\psi^{\pm}$ on $x$ is changed by the presence of $\rho(x)$. 
The last two equations of (\ref{pchobr}) tell us simply that $u(k)$ plays a role of a morphism between the representations $\rho(R)$ and $D^j(R)$, whereas $v(k)$ is a morphism between $\rho(R)$ and $D^{j*}(R)$. The representations $D^j$ (and $D^{j*}$) are irreducible. If $\rho$ provided an irreducible representation as well when restricted to the rotational subgroup of the Poincar{\'e} group, then Shur's lemma would imply that the amplitudes either vanish or are isomorphisms (i.e. square matrices). In general, however, $\rho(R)$ is not irreducible, but rather acquires (in the appropriate basis) a block--diagonal form
\be
\rho(R)=
\left({
\begin{array}{cccc}
\rho^1(R) & 0 & 0 \\
0 & \ddots & 0 \\
0 & 0 & \rho^M(R)
\end{array}
}\right)\quad ,
\ee
where the representations $\rho^i$ are irreducible and zeros mean zero matrices of appropriate shapes and dimensions. If the amplitudes are divided correspondingly as
\be\label{uivi}
u(k)=\left({
\begin{array}{ccc}
u^1 \\
\vdots \\
u^M
\end{array}
}\right)\quad,\quad
v(k)=\left({
\begin{array}{ccc}
v^1 \\
\vdots \\
v^M
\end{array}
}\right)\quad,
\ee
where the number of rows of the matrices $u^i$, $v^i$ equals the dimension of the corresponding representation $\rho^i$, then the last two equations of(\ref{pchobr}) will reduce to the collection of matrix equations $\rho^i(R)u^i=u^iD^j(R)$, $\rho^i(R)v^i=v^iD^{j*}(R)$, $i=1,\dots M$. All the representations that occur here are now irreducible and hence Shur's lemma applies. For a fixed value of spin $j$ it follows that $u^i$ (or $v^i$) can be nonzero only if the corresponding representation $\rho^i$ has dimension $2j+1=dim\l{D^j}\r=dim\l{D^{j*}}\r$. Hence, the representation $\rho$ can describe a particle with spin $j$ only if it contains at least one $2j+1$--dimensional irreducible representation of the group of rotations. After the last two equations of (\ref{pchobr}) are solved, the amplitudes are composed of blocks of $(2j+1)\times (2j+1)$--dimensional non--zero matrices, possibly separated by some blocks of zeros. This zeros may seem superfluous at first, but they may be filled by non--zero expressions when the amplitudes are busted (e.g. for Lorentz--vector field of spin $j=1$). Also, the different non--zero blocks that transform completely independently under rotations may be mixed by discrete symmetries such as parity (e.g. the Dirac field) or by the action of the complete representation $\rho$ of the Poincar{\'e} group (e.g. Poincar{\'e}--vector field, see Section \ref{Poinv}).

In order to satisfy requirements of conservation of electric charge and self--adjointness of an interaction density composed of causal fields, it is necessary to consider the combinations \cite{Wein}
\be\label{psitot}
\psi(x)=\psi^{+}(x)+{\psi^{-}}^c(x)\quad, 
\ee 
where ${\psi^{-}}^c(x)=\int d\Gamma_p v(x,p){a^c}^{\dag}_{p}$ and $c$ stands for anti--particle (for electrically neutral particles $a^c=a$).
The fields $\psi^{\pm}$ and ${\psi^{\pm}}^c$, and hence also $\psi$, 
are assumed to transform according to the same representation $\rho$ of the Poincar{\'e} group.
The fields should also satisfy the causality condition -- the commutator
\be\label{przy}
\ba
&[\psi_l(x),\psi^{\dag}_{l'}(x')]_{\mp}=\left[{\rho(x)\int\l{
e^{-ip\cdot (x-x')}N(p)\mp
e^{ip\cdot (x-x')}M(p)}\r d\Gamma_p \ \rho^{\dag}(x')}\right]_{ll'}, \\
&N(p)=u(p)u^{\dag}(p), \quad M(p)=v(p)v^{\dag}(p),
\ea
\ee
ought to vanish for space--like interval $x-x'$ (also
$[\psi_l(x),\psi_{l'}(x')]_{\mp}$ should satisfy the condition, but 
for charged particles it is fulfilled automatically). 

\vskip 0.7 in
\noindent{\bf Discrete symmetries}

If there are many non--zero blocks $u^i$ or $v^i$ in the decomposition (\ref{uivi}) of the amplitudes, then the relative weights of these blocks will not be fixed by equations (\ref{pchobr}). One can make use of discrete symmetries $C, P$ and $T$ to limit this arbitrariness. The parity $P$ appears to be particularly important. Let $\mathcal{P}=diag(1,-1,-1,-1)$ represent the parity operation in Minkowski space. In order for the quantum theory to be parity invariant, there should exist a unitary transformation ${\mathrm P}$ acting on the space of states and satisfying the following commutation relations with the generators of the unitary representation $U$ of the Poincar{\'e} group
\be\label{parcom}
\mathrm{P}P^a\mathrm{P}^{-1}={\mathcal{P}_b}^aP^b\quad,
\quad \mathrm{P}J^{ab}\mathrm{P}^{-1}={\mathcal{P}_c}^a{\mathcal{P}_d}^bJ^{cd}\quad.
\ee
The action of this transformation on annihilation and creation operators is
\be
\mathrm{P}a_{p,\sigma}\mathrm{P}^{-1}=\eta^*a_{\mathcal{P}p,\sigma}\quad,\quad
\mathrm{P}{a^c}^{\dag}_{p,\sigma}\mathrm{P}^{-1}=\eta^c{a^c}^{\dag}_{\mathcal{P}p,\sigma}\quad,
\ee
where $\eta$ ($\eta^c$) is the internal parity of the particle (anti--particle). The action of parity on fields is then 
\be\label{parpsi}
\ba
&\mathrm{P}\psi^+(x)\mathrm{P}^{-1}=\eta^*\rho(x)\int
e^{-ip\cdot\mathcal{P}x}\rho\l{L_{\mathcal{P}p}}\r
u(k)a_p \, d\Gamma_p\quad,\\
&\mathrm{P}{\psi^-}^c(x)\mathrm{P}^{-1}=\eta^c\rho(x)\int
e^{ip\cdot\mathcal{P}x}\rho\l{L_{\mathcal{P}p}}\r
v(k){a^c}^{\dag}_p \, d\Gamma_p\quad.
\ea
\ee
We have suppressed the indices $l$ and $\sigma$ (think of the above formulas in terms of matrix multiplication). Also the change of integration variables was performed $p\rightarrow\mathcal{P}p$ and the invariance of the measure was employed.
The transformation formula may acquire a simple form when expressed in terms of causal fields if there exists a matrix 
$\stackrel{\rho}{\mathcal{P}}$, acting in the linear space of representation $\rho$, such that
\be\label{parwar1}
\ba
&\rho\l{L_{\mathcal{P}p}}\r=\stackrel{\rho}{\mathcal{P}}\rho(L_p)\stackrel{\rho}{\mathcal{P}}\quad,\\
&\stackrel{\rho}{\mathcal{P}}u(k)=b_uu(k)\quad,\quad \stackrel{\rho}{\mathcal{P}}v(k)=b_vv(k)\quad,\quad b_u,b_v\in\mathbb{C}\quad.
\ea
\ee
Then (\ref{parpsi}) is reduced to
\be\label{parwar2}
\ba
&\mathrm{P}\psi^+(x)\mathrm{P}^{-1}=\eta^*b_u\rho(x)\stackrel{\rho}{\mathcal{P}}\rho^{-1}(\mathcal{P}x)
\ \psi^+(\mathcal{P}x)\quad,\\
&\mathrm{P}{\psi^-}^c(x)\mathrm{P}^{-1}=\eta^cb_v\rho(x)\stackrel{\rho}{\mathcal{P}}\rho^{-1}(\mathcal{P}x)
\ {\psi^-}^c(\mathcal{P}x)\quad.
\ea
\ee
It is clear that in order for the total field (\ref{psitot}) to transform reasonably under parity, the relation $\eta^cb_v=\eta^*b_u$ must hold. Then the field transforms according to
\be
\mathrm{P}\psi(x)\mathrm{P}^{-1}=\eta^*b_u\rho(x)\stackrel{\rho}{\mathcal{P}}\rho^{-1}(\mathcal{P}x)
\ \psi(\mathcal{P}x)\quad.
\ee
Note that the transformation in general explicitly depends on $x$. This feature of the theory may seem strange at first, but in fact does not lead to any detectable physical effects since the fields themselves are not directly measurable. Rather the interaction terms obtained from them are physically important and these are parity invarant, if constructed properly.

The remaining discrete symmetries can be similarly implemented by introduction of a unitary operator $C$ and anti--unitary operator $T$ that act on the annihilation and creation operators as
\be
Ca^{\dag}_{p,\sigma}C^{-1}=\xi a^{\dag}_{p,\sigma}\quad,\quad Ta^{\dag}_{p,\sigma}T^{-1}=\zeta (-1)^{j-\sigma} a^{\dag}_{\mathcal{P}p,-\sigma}\quad,
\ee
where $\xi$ and $\zeta$ are the internal phases of charge conjugation and time reversal. For general representation $\rho$ of the Poincar\'e group, their action on fields will exhibit explicite dependence on the space--time point, just as in the case of parity.

Let us now consider the examples. The only one--dimensional representation of the Poincar{\'e} group is the trivial one, therefore there is no need to consider scalar field. For the vector field the relevant representation of the Lorentz group is its fundamental representation in $\mathbb{R}^4$, $\rho(\Lambda){^a}_b=\Lambda{^a}_b$, which cannot be extended to the faithful representation of the whole Poincar{\'e} group in $\mathbb{R}^4$. The situation is much different for the Dirac field and the Poincar{\'e}--vector field. We shall consider these two cases separately.

\subsection{The Dirac field}\label{Dirac}

The standard spinor representation of the Lorentz group is given by the generators 
\be\label{JJ}
\mathbb{J}^{ab}=-\frac{i}{4}[\gamma^a,\gamma^b]_{-}\quad,
\ee
 where $\gamma^a$ are the Dirac matrices satisfying $[\gamma^a,\gamma^b]_{+}=2\eta^{ab}{\bf 1}$,
for which we shall choose a convenient representation
\be\label{dir}
\ba
\gamma^a=
\l
\begin{array}{cc}
0 & \sigma^a \\
\bar{\sigma}^a & 0 \\
\end{array}
\r, \quad
\sigma^0=\bar{\sigma}^0={\bf 1},\quad \bar{\sigma}^i=-\sigma^i,
\ea
\ee
where $\sigma^i$ are Pauli matrices. 
The representation admits a unique extension to the faithful representation of the Poincar{\'e} group on $\mathbb{C}^4$ provided by the generators of translations
\be\label{PP}
\mathbb{P}^a=\alpha \gamma^a(1+s\gamma^5)\quad,
\ee
where $\gamma^5=-i\gamma^0\gamma^1\gamma^2\gamma^3$, $s=\pm 1$ and $\alpha$ is a parameter of dimension of mass in natural units $c=\hbar=1$. We shall restrict ourselves to real values of $\alpha$ for which the representation $\rho$ satisfies a pseudo--unitarity condition $\rho^{\dag}(\Lambda,b)=\gamma^0\rho^{-1}(\Lambda,b)\gamma^0$.
The conditions (\ref{pchobr}), which can be imposed on the amplitudes for $j=1/2$ only because of Shur's lemma, lead to the following form 
of the amplitudes for standard momentum
\be\label{ukvk}
u(k)=
\l
\begin{array}{cccc}
c_+ & 0 \\
0 & c_+ \\
c_- & 0 \\
0 & c_-
\end{array}
\r\quad, \quad
v(k)=
\l
\begin{array}{cccc}
0 & -d_+ \\
d_+ & 0 \\
0 & -d_- \\
d_- & 0
\end{array}
\r \quad,\quad
c_+, c_-, d_+, d_- \in \mathbb{C}.
\ee
One can then calculate $u(p)$ and $v(p)$ using
\be\label{DLp}
\rho\l{L_p}\r=\rho^{\dag}\l{L_p}\r=\frac{m+p_a\gamma^a\gamma^0}{\sqrt{2m\l{p^0+m}\r}}\quad.
\ee
It can now be readily proved that the causality condition (\ref{przy}) will be fulfilled if and only if
\be\label{warprzycz}
c_+c_-^*=\pm d_+d_-^* \quad ,\quad
|c_+|^2=\mp |d_+|^2\quad,\quad
|c_-|^2=\mp |d_-|^2 \quad.
\ee
The last two equations can be satisfied only with the lower sign. It follows that the modified Dirac field necessarily describes fermions of spin $\frac{1}{2}$, just like the standard one. 

Let us finally investigate whether the theory can be made manifestly parity invariant. 
The relations (\ref{parwar1}) are valid for $\stackrel{\rho}{\mathcal{P}}=\gamma^0$. Since $\gamma^0\gamma^0={\mathbf 1}_4$, it is necessary that $b_u,b_v=\pm 1$.
It then follows from (\ref{parwar1}), (\ref{ukvk}) and (\ref{warprzycz}) that $b_v=-b_u$, $|c_-|=|c_+|=|d_-|=|d_+|$, $c_-=b_uc_+, d_-=-b_ud_+$. If $\psi$ is to have well established transformation properties with respect to parity, it is necessary that $\eta^c=-\eta^*$. Finally, using the possibility of changing relative phase of annihilation and creation operators, the freedom of performing global rescaling of the field and that of replacing $\psi$ by $\gamma^5\psi$, we can cast the amplitudes into the standard form
\be\label{ukvk1}
u(k)=\sqrt{m}
\l
\begin{array}{cccc}
1 & 0 \\
0 & 1 \\
1 & 0 \\
0 & 1
\end{array}
\r, \quad
v(k)=\sqrt{m}
\l
\begin{array}{cccc}
0 & -1 \\
1 & 0 \\
0 & 1 \\
-1 & 0
\end{array}
\r\quad,
\ee
where the factor $\sqrt{m}$ is necessary in order for $\psi$ to have dimension $-3/2$.
Note that the action of parity transformation on the field $\psi(x)$ explicitly depends on $x$, 
${\mathrm P}\psi(x){\mathrm P}^{-1}=\eta^*\rho(x)\gamma^0\rho^{-1}(\mathcal{P}x)\psi(\mathcal{P}x)$.
This dependence would not be present if the generators $\mathbb{P}$
satisfied the appropriate commutation relations with $\gamma^0$
\be\label{pbbcom}
\gamma^0\mathbb{P}^a\gamma^0={\mathcal{P}_b}^a\mathbb{P}^b
\ee
(compare (\ref{parcom})). However, the only possible nontrivial generators for the Dirac field (\ref{PP}) do not possess this property. The $C$ and $T$ symmetries can be implemented without further restrictions on the amplitudes.

Since the field is of the form 
\be\label{rel}
\psi(x)=\rho(x)\tilde{\psi}(x)\quad,
\ee
where $\tilde{\psi}(x)$ possesses all the properties of the standard Dirac field, the modified field $\psi$ should satisfy the equation derived from (\ref{rel}) under the assumption that $\tilde{\psi}$ obeys the usual Dirac equation. Explicitly,
\be\label{direq}
\ba
&\l{i\gamma^a\partial_a-m}\r\tilde{\psi}(x)=0\quad\Rightarrow\quad 
\left[{\tilde{\gamma^a}(x)\l{i\partial_a+\mathbb{P}_a}\r-m}\right]\psi(x)=0\quad,\\
&\tilde{\gamma^a}(x):=\rho(x)\gamma^a\rho^{-1}(x)\quad.
\ea
\ee
This equation can be derived from the Lagrangian density
\be\label{modL0}
\mathcal{L}_0=\ov{\psi}(x)\left[{\tilde{\gamma^a}(x)\l{i\partial_a+\mathbb{P}_a}\r-m}\right]\psi(x)\quad,
\ee
or the Lagrangian four--form
\be
\mathfrak{L}_0=-i\l{\star dx_a}\r\wedge \ov{\psi}\tilde{\gamma^a}d\psi-\ov{\psi}\l{m-\tilde{\gamma}^a\mathbb{P}_a}\r\psi\,d^4x\quad,
\ee
where $\ov{\psi}=\psi^{\dag}\gamma^0$ is the Dirac conjugation, $\star$ is the Hodge star of Minkowski metric, i.e. $\star dx_a=\frac{1}{6}\epsilon_{abcd}dx^b\wedge dx^c\wedge dx^d$, where $\epsilon_{abcd}$ is the totally anti--symmetric symbol with $\epsilon_{0123}=1$, and $d^4x=dx^0\wedge dx^1\wedge dx^2\wedge dx^3$ is the volume form of Minkowski metric. This four--form is clearly Poincar{\'e} invariant under the global action of the relevant representation,
\be
\psi\rightarrow \rho(\Lambda,b)\psi \quad \Rightarrow\quad \ov{\psi}\rightarrow\ov{\psi}\rho^{-1}(\Lambda,b)\quad,\quad 
\tilde{\gamma}^a \rightarrow {\Lambda^a}_b\rho(\Lambda,b)\tilde{\gamma}^b\rho^{-1}(\Lambda,b)\quad,\quad dx^a\rightarrow{\Lambda^a}_bdx^b\quad. 
\ee
The transformation formula for $\tilde{\gamma^a}$ follows from 
\be
\ba
&\tilde{\gamma}^a(\Lambda x+b)\,=\,\rho(\Lambda x+b)\gamma^a\rho^{-1}(\Lambda x+b)\,=\,\rho(\Lambda,b)\rho(x)\rho^{-1}(\Lambda)\gamma^a\rho(\Lambda)\rho^{-1}(x)\rho^{-1}(\Lambda,b)\\
&={\Lambda^a}_b\rho(\Lambda,b)\tilde{\gamma}^b(x)\rho^{-1}(\Lambda,b)\quad.
\ea
\ee

The fact that the new field is related via (\ref{rel}) to the standard Dirac field seems to suggest that all the physical properties of $\psi$ will be indistinguishable from those of $\tilde{\psi}$. This supposition is further supported by observation that all the Noether currents of physical importance will express in exactly the same way in terms of annihilation and creation operators when calculated for the field $\tilde{\psi}$ and $\psi$ (see the Appendix \ref{A2} for the proof). However, considering the generalized field that transforms under the faithful representation of the Poincar{\'e} group has important consequences for PGT that will be discussed in the forthcoming paper in which gravity will be included.

\subsection{The Poincar{\'e}--vector field}\label{Poinv}

Let us now consider a faithful representation of the Poincar{\'e} group in $\mathbb{R}^5$ defined by
\be\label{PVrepr}
\rho(\Lambda,b)=
\l
\begin{array}{cc}
\Lambda & \alpha b \\
0 & 1 
\end{array}
\r\quad,\quad
\Lambda\in SO(1,3)\quad,\quad b\in\mathbb{R}^4\quad,\quad \alpha\in\mathbb{R}\quad.
\ee 
The parameter $\alpha$ corresponds to the possibility of rescaling of $\mathbb{P}$ and thus is analogues to $\alpha$ that was introduced for the Dirac field.
The representation of the group of rotations that is contained in $\rho$ is a simple sum of two trivial representations and the fundamental one. The corresponding matrix acquires a block--diagonal form
\be
\rho(R)=
\l
\begin{array}{ccc}
1 & 0 & 0 \\
0 & R & 0 \\
0 & 0 & 1 
\end{array}
\r\quad,\quad
R\in SO(3)\quad,
\ee 
where zeros are zero--matrices of appropriate shapes and dimensions. Using the notation introduced in (\ref{uivi}) one can conclude that the last two equations of (\ref{pchobr}) can be solved either for $j=1,\, u^1=u^3=v^1=v^3=0$ or for 
$j=0,\,u^2=v^2=0$. 

\vskip 0.1 in
\noindent{\bf The $j=1$ case} 

This case is not really interesting, since the amplitude is then of the form
\be
u(p)=\rho(L_p)u(k)=
\l
\begin{array}{cc}
L_p & 0 \\
0   & 1  
\end{array}
\r
\l
\begin{array}{cc}
\tilde{u}(k) \\
0  
\end{array}
\r=
\l
\begin{array}{cc}
\tilde{u}(p) \\
0  
\end{array}
\r
\quad,
\ee 
where $\tilde{u}$ is the amplitude for the standard vector field. From (\ref{uxvx}) it then follows that the total $x$--dependent amplitude is of the form 
\be
u(x,p)=e^{-ip\cdot x}\rho(x)u(p)=e^{-ip\cdot x}
\l
\begin{array}{cc}
{\bf 1}_4 & \alpha x \\
0   & 1  
\end{array}
\r
\l
\begin{array}{cc}
\tilde{u}(p) \\
0  
\end{array}
\r=e^{-ip\cdot x}u(p)
\quad.
\ee
Similar result holds for $v$. The creation and annihilation fields are thus Fourier transforms of the standard momentum--dependent amplitudes for vector field of spin 1, with the unimportant row of zeros added.

\vskip 0.1 in
\noindent{\bf The j=0 case}

Since the amplitudes are of the form
\be
u(k)=
\l
\begin{array}{cc}
\frac{c_0}{m}\, k \\
c_4  
\end{array}
\r\quad,\quad
v(k)=
\l
\begin{array}{cc}
\frac{d_0}{m}\, k \\
d_4
\end{array}
\r
\quad,\quad c_0,c_4,d_0,d_4\in\mathbb{C}\quad,\quad 
\ee
where $k=(m,0,0,0)$ is the standard momentum, it follows that
\be
\ba
&u(x,p)=e^{-ip\cdot x}\rho(L_p,x)u(k)=e^{-ip\cdot x}
\l
\begin{array}{cc}
\frac{c_0}{m}\, p+c_4\alpha x \\
c_4  
\end{array}
\r\quad,\\
&v(x,p)=e^{ip\cdot x}\rho(L_p,x)v(k)=e^{ip\cdot x}
\l
\begin{array}{cc}
\frac{d_0}{m}\, p+d_4\alpha x \\
d_4  
\end{array}
\r\quad
\ea
\ee
(think of $x$ and $p$ as column matrices whose entries are the components of four--momentum and Minkowskian coordinates, respectively).

Note that parity invariance does not limit the freedom of choice of the parameters at all. The relations (\ref{parwar1}) are satisfied for
\be
\stackrel{\rho}{\mathcal{P}}=
\l
\begin{array}{cc}
\mathcal{P} & 0 \\
0 & 1
\end{array}
\r\quad,\quad b_u=b_v=1\quad.
\ee
What is more, the parity transformation acts on fields in an $x$--independent way, since 
$\rho(x)\stackrel{\rho}{\mathcal{P}}\rho^{-1}(\mathcal{P}x)=\stackrel{\rho}{\mathcal{P}}$. 
The other discrete symmetries $C$ and $T$ can also be easily implemented without any further restrictions on the parameters.

The causality condition (\ref{warprzycz}) is satisfied if and only if
\be\label{caus}
\ba
&|c_0|^2(\partial_a\partial_b\triangle)(x-x')\mp |d_0|^2(\partial_a\partial_b\triangle)(x'-x)=0\quad,\\
&c_0c_4^*(\partial_a\triangle)(x-x')\mp d_0d_4^*(\partial_a\triangle)(x'-x)=0\quad,\\
&c_0^*c_4(\partial_a\triangle)(x-x')\mp d_0^*d_4(\partial_a\triangle)(x'-x)=0\quad,\\
&|c_4|^2\triangle(x-x')\mp |d_4|^2\triangle(x'-x)=0\quad
\ea
\ee
for space--like $x-x'$, where $\triangle(x):=\int e^{-ip\cdot x}d\Gamma_p$. The function $\triangle$ is even for space--like $x$, hence its derivative is odd and the second derivative is again even. Hence, the condition reduces to
\be
|c_0|^2\mp |d_0|^2=0\quad,\quad 
|c_4|^2\mp |d_4|^2=0\quad,\quad 
c_0c_4^*\pm d_0d_4^*=0\quad,\quad
\ee
that can be satisfied with the upper sign only. Hence, the particles under investigation are bosons. Adjusting the relative phase of the annihilation and creation operators and rescaling globally the field $\psi=\psi^++\psi^{-c}$ it is possible to achieve $c_4=d_4=1$ (note that the dimension of the field was thus determined to be $-1$, as it should for a spinless particle). After this is done, the phases and scaling are fixed, so one cannot perform the same operation on $c_0$ and $d_0$. However, from (\ref{caus}) it now follows that $d_0=-c_0$. The remaining freedom of the parameters $c_0$ and $\alpha$ cannot be restricted by discrete symmetries. We shall see however below that the Lagrangian formulation suggests a particular value of $c_0$. 

After the causality condition is imposed on the amplitudes, the field is equal to
\be\label{PVF}
\ba
&\psi(x)=\int \l{u(x,p)a_p+v(x,p)a^{c\dag}_p}\r\,d\Gamma_p=\rho(x)\tilde{\psi}(x)=
\l
\begin{array}{cc}
{\tilde{\Phi}}(x)+\alpha x\phi(x) \\
\phi(x)
\end{array}
\r=
\l
\begin{array}{c}
 \Phi(x) \\
\phi(x)
\end{array}
\r
\quad,\\
&\phi(x)=\int\l{e^{-ip\cdot x}a_p+e^{ip\cdot x}a^{c\dag}_p}\r\,d\Gamma_p\quad,\quad 
{\tilde{\Phi}}^a(x)=\frac{ic_0}{m}\l{\partial^a\phi}\r(x)\quad,\quad
\Phi(x)={\tilde{\Phi}}(x)+\alpha x\phi(x)\quad,
\ea
\ee
where
\be\label{tPVF}
\tilde{\psi}=
\l
\begin{array}{cc}
{\tilde{\Phi}} \\
\phi
\end{array}
\r
\ee
satisfies the Klein--Gordon equation $\l{\square+m^2}\r\tilde{\psi}=0$. Similarly to the Dirac field case, note that $m$ is just the parameter that determines the mass shall ($p\cdot p=m^2$), and hence the invariant measure $d\Gamma_p$, and has nothing to do with $\alpha$, the latter being related to the way of embedding the Poincar{\'e} group 
in $End(\mathbb{R}^5)$. 

The relevant field equations, when written in terms of the components of $\tilde{\psi}$, are
\be
{\tilde{\Phi}}^a=\frac{ic_0}{m}\,\partial^a\phi\quad,\quad \l{\square+m^2}\r\phi=0\quad.
\ee
To provide a Lagrangian formulation for them, we shall consider separately the case of real and complex field. If $\psi$ is to be a real field, it is necessary that $c_0=ir$ for some $r\in\mathbb{R}$. The Lagrangian can then be given as
\be\label{PVL}
\mathcal{L}=-\frac{m}{r}{\tilde{\Phi}}^a\partial_a\phi-\frac{1}{2}\l{\frac{m}{r}}\r^2{\tilde{\Phi}}_a{\tilde{\Phi}}^a-\frac{1}{2}m^2\phi^2\quad.
\ee
It would be useful to be able to express the Lagrangian in terms of the entire field $\tilde{\psi}$ and matrices representing operations of well established physical or geometrical meaning. It appears that this is possible only if $r=-s$, where $s=\pm 1$. Then
\be\label{PVLpsit}
\mathcal{L}=\frac{sm}{\alpha}\tilde{\psi}^T\stackrel{\rho}{\mathcal{P}}\l{i\mathbb{P}^a}\r\partial_{a}\tilde{\psi}-\frac{1}{2}m^2{\tilde{\psi}}^{T}\stackrel{\rho}{\mathcal{P}}\tilde{\psi}
\quad.
\ee 
Certainly, one could express (\ref{PVL}) in terms of $\tilde{\psi}$ for any value of $r$ using the matrices
$
\l
\begin{array}{cc}
\eta & 0 \\
0 & 0
\end{array}
\r
$
and
$
\l
\begin{array}{cc}
0 & 0 \\
0 & 1
\end{array}
\r
$
but these do not posses such a straightforward physical interpretation as $\stackrel{\rho}{\mathcal{P}}$, which simply represents the implementation of parity transformation on the space of fields. Therefore, we will further consider the Lagrangian density (\ref{PVLpsit}), which gives rise to the following Lagrangian density for $\psi$
\be
\ba
&\mathcal{L}\l{\psi,\partial_a\psi,x}\r=\frac{sm}{\alpha}\psi^T\tilde{\mathcal{P}}(x)\mathbb{P}^a\l{i\partial_{a}+\mathbb{P}_a}\r\psi-\frac{1}{2}m^2\psi^{T}\tilde{\mathcal{P}}(x)\psi
\quad,\\
&\tilde{\mathcal{P}}(x):=\rho^T(-x)\stackrel{\rho}{\mathcal{P}}\rho(-x)\quad.
\ea
\ee
Note that the entries of the matrices $\mathbb{P}^a$ are imaginary and hence the Lagrangian is real. For later convenience, let us rewrite this Lagrangian density as a Lagrangian four--form
\be\label{PVlff}
\mathfrak{L}=-\frac{smi}{\alpha}\l{\star dx_a}\r\wedge\psi^T\tilde{\mathcal{P}}(x)\mathbb{P}^ad\psi-
\frac{1}{2}m^2\psi^T\tilde{\mathcal{P}}(x)\psi d^4x\quad,
\ee
where we have used the fact that $\mathbb{P}^a\mathbb{P}^b=0$ for the representation under consideration. This four--form is clearly invariant under the global action of the Poincar\'e group $\psi\rightarrow\rho(\Lambda,b)\psi$, $x\rightarrow \Lambda x+b$. To see this, use
$\tilde{\mathcal{P}}\rightarrow {\rho^{-1}}^T(\Lambda,b)\tilde{\mathcal{P}}\rho^{-1}(\Lambda,b)$ and $\rho^{-1}(\Lambda,b)\mathbb{P}^a\rho(\Lambda,b)={\Lambda^a}_c\mathbb{P}^c$ 
(the first transformation formula follows from $\tilde{\mathcal{P}}(\Lambda x+b)={\rho^{-1}}^T(\Lambda,b)\tilde{\mathcal{P}}(x)\rho^{-1}(\Lambda,b)$ and the second from the commutation relations for the Poincar\'e algebra).

If the field is complex, $c_0$ need not be restricted to the imaginary values and the real Lagrangian density generating the appropriate field equations is 
\be\label{PVparts}
\mathcal{L}=\frac{im}{c_0^*}{\tilde{\Phi}}^{a*}\partial_a\phi-\frac{im}{c_0}{\tilde{\Phi}}^a\partial_a\phi^*-\l{\frac{m}{|c_0|}}\r^2{\tilde{\Phi}}_a^*{\tilde{\Phi}}^a-m^2\phi^*\phi\quad.
\ee
To express it in terms of $\tilde{\psi}$, one has to assume that $c_0$ is a pure phase, $c_0=e^{i\beta}$, $\beta\in\mathbb{R}$. The Lagrangian is then
\be\label{PVch}
\mathcal{L}=-\frac{m}{\alpha}\l{e^{i\beta}{\tilde{\psi}}^{\dag}\stackrel{\rho}{\mathcal{P}}\mathbb{P}^a\partial_a\tilde{\psi}+e^{-i\beta}\partial_a{\tilde{\psi}}^{\dag}{\mathbb{P}^a}^{\dag}\stackrel{\rho}{\mathcal{P}}\tilde{\psi}}\r-m^2{\tilde{\psi}}^{\dag}\stackrel{\rho}{\mathcal{P}}\tilde{\psi}\quad.
\ee
The Lagrangian density (and four--form) for $\psi$ can be found by using $\tilde{\psi}(x)=\rho^{-1}(x)\psi(x)$. The important difference when compared to the real case is that the free parameter $c_0$ was fixed by Lagrangian formalism only up to the choice of real parameter $\beta$. Note that the variation with respect to ${\tilde{\Phi}}^a$ in (\ref{PVparts}) (or with respect to the first four components of ${\tilde{\psi}}^{\dag}$ in (\ref{PVch})) gives the equation ${\tilde{\Phi}}^a=\frac{ic_0}{m}\partial^a\phi$ (or ${\tilde{\Phi}}^a=\frac{i}{m}e^{i\beta}\partial^a\phi$), which, when inserted back to (\ref{PVparts}) (or (\ref{PVch})), leads to the standard Lagrangian density for a complex scalar field
\be\label{Lst}
\mathcal{L}_{st}=\partial_a\phi^*\partial^a\phi-m^2\phi^*\phi\quad.
\ee

\section{Interactions}\label{Int}

\subsection{General procedure for constructing interaction terms}\label{intgen}

In order to assert the Lorentz invariance of the $S$ matrix, conform to the cluster decomposition principle and ensure electric charge conservation in standard quantum field theory \cite{Wein}, the interaction needs to be described in terms of the interaction density $\mathcal{H}(x)$ constructed from fields according to
\be
\mathcal{H}(x)=\sum_N \sum_{l_1\dots l_N} g_{l_1\cdots l_N} :\psi^{(1)}_{l_1}(x)\cdots \psi^{(N)}_{l_N}(x): \quad,
\ee 
where $\psi^{(i)}$ are causal fields constructed according to all the principles reviewed in Section \ref{free}, $:\ :$ denotes normal ordering and $g_{l_1\cdots l_N}$ are numerical coefficients that satisfy
\be\label{gcond}
\sum_{l_1\dots l_N} g_{l_1\cdots l_N}\rho^{(1)}\l{\Lambda^{-1}}\r_{l_1l_1'}\cdots \rho^{(N)}\l{\Lambda^{-1}}\r_{l_Nl_N'}=g_{l_1'\cdots l_N'} \quad.
\ee
if the fields $\psi^{(i)}$ transform trivially under translations, this procedure guaranties that the interaction density is a scalar, $U_0(\Lambda,b)\mathcal{H}(x)U_0^{-1}(\Lambda,b)=\mathcal{H}(\Lambda x+b)$
\footnote{
The subscript $0$ in $U_0$ means that the generators of the transformation $U_0(\Lambda,b)$ are those of free theory and not the interacting one. Also the fields that appear in this subsection are the interaction picture fields whose evolution is governed by the Hamiltonian of the free theory.
}. If, however, the fields do transform under faithful representations of the Poincar\'e group, then the procedure should be modified. The coefficients $g_{l_1\cdots l_N}$ have to be replaced by functions,
\be
g_{l_1\cdots l_N}(x)=\tilde{g}_{k_1\cdots k_N}\rho^{(1)}_{k_1l_1}(-x)\cdots\rho^{(N)}_{k_Nl_N}(-x)\quad,\qquad \rho^{(i)}(x)=e^{ix\cdot\mathbb{P}^{(i)}}\quad,
\ee
where ${\mathbb{P}^{(i)}}^a$ are the generators of translations of the representation of the Poincar\'e group under which $\psi^{(i)}$ transforms and the numerical coefficients $\tilde{g}_{l_1\cdots l_N}$ satisfy the standard condition (\ref{gcond}). This generalized procedure guarantees the scalar nature of $\mathcal{H}$ in the case of nontrivial realization of translations. However, when it is applied, the interaction density $\mathcal{H}$ appears to be the same as the one that would be obtained in a standard way from the Lorentz transforming fields $\tilde{\psi}^{(i)}(x)={\rho^{(i)}}^{-1}(x)\psi^{(i)}(x)$ with the coefficients $\tilde{g}_{l_1\cdots l_N}$. This observation shows very clearly that the scattering theory of the fields $\psi^{(i)}$ that transform non-trivially under translations is necessarily equivalent to the theory of the fields $\tilde{\psi}^{(i)}$ that do not feel translations.

Surprisingly enough, in spite of the above mantioned equivalence, considering fields that transform under faithful representations of the Poincar\'e group may still lead to the physically interesting consequences. Although one can always find a theory of a Lorentz field that is equivalent to a given theory of the Poincar\'e field, this Lorentz field theory may seem to be an  artificial one in such a way that normally it would not be considered at all. Only the recognition that it is equivalent to a theory of some indecomposable\footnote{
The notion of indecomposability should not be confused with that of irreducibility. See e.g. \cite{Hall} for definitions.
}  representation of the Poincar\'e group can make it worth consideration. To illustrate this issue, let us consider a spin zero real field, equipped with the simplest potential term that yields a renormalizable theory with positive--definite Hamiltonian. If the standard theory of a scalar field is used, the relevant potential term is $\lambda\phi^4$, $\lambda>0$. This term is just proportional to the square of a mass term in the free Lagrangian. If the Poincar\'e--vector field is used instead, constructing a potential proportional to the square of a ``mass term'' (the term proportional to $m^2$ in (\ref{PVLpsit})) would result in a theory with the Lagrangian density
\be
\mathcal{L}=\frac{sm}{\alpha}\tilde{\psi}^T\stackrel{\rho}{\mathcal{P}}\l{i\mathbb{P}^a}\r\partial_{a}\tilde{\psi}-\frac{1}{2}m^2{\tilde{\psi}}^{T}\stackrel{\rho}{\mathcal{P}}\tilde{\psi}-\lambda\l{{\tilde{\psi}}^{T}\stackrel{\rho}{\mathcal{P}}\tilde{\psi}}\r^2\quad,
\ee 
which would result in non--positive Hamiltonian and hence should be rejected. There is, however, another natural candidate for a potential, namely the term
$\tilde{\psi}^T\stackrel{\rho}{\mathcal{P}}\l{\frac{i}{\alpha}\mathbb{P}^a}\r\tilde{\psi}
\tilde{\psi}^T\stackrel{\rho}{\mathcal{P}}\l{\frac{i}{\alpha}\mathbb{P}_a}\r\tilde{\psi}
=\phi^2\tilde{\Phi}_a\tilde{\Phi}^a$. The Lagrangian
\be\label{PVLpot}
\ba
\mathcal{L}&=\frac{sm}{\alpha}\tilde{\psi}^T\stackrel{\rho}{\mathcal{P}}\l{i\mathbb{P}^a}\r\partial_{a}\tilde{\psi}-\frac{1}{2}m^2{\tilde{\psi}}^{T}\stackrel{\rho}{\mathcal{P}}\tilde{\psi}-\frac{\lambda}{2}\tilde{\psi}^T\stackrel{\rho}{\mathcal{P}}\l{\frac{i}{\alpha}\mathbb{P}^a}\r\tilde{\psi}
\tilde{\psi}^T\stackrel{\rho}{\mathcal{P}}\l{\frac{i}{\alpha}\mathbb{P}_a}\r\tilde{\psi} \\
&=sm\tilde{\Phi}^a\partial_a\phi-\frac{1}{2}m^2\tilde{\Phi}_a\tilde{\Phi}^a-\frac{1}{2}m^2\phi^2-\frac{\lambda}{2}\phi^2\tilde{\Phi}_a\tilde{\Phi}^a
\ea
\ee 
yields the Hamiltonian density
\be
t_{00}=\frac{\partial\mathcal{L}}{\partial(\partial_0\phi)}\partial_0\phi-\mathcal{L}=\frac{1}{2}m^2\left[{\l{\tilde{\Phi}^0}\r^2+\phi^2}\right]-sm\tilde{\Phi}^i\partial_i\phi-\frac{1}{2}m^2\tilde{\Phi}^i\tilde{\Phi}^i+\frac{1}{2}\lambda\phi^2\left[{\l{\tilde{\Phi}^0}\r^2-\tilde{\Phi}^i\tilde{\Phi}^i}\right]
\ee
(here $t_{00}$ is the relevant component of the Noether energy--momentum tensor). It is not clear from this expression that the Hamiltonian is positive. However, if the field equation  
\be
sm\partial_a\phi=m^2\tilde{\Phi}_a+\lambda\phi^2\tilde{\Phi}_a\quad
\ee
obtained form the variation of (\ref{PVLpot}) with respect to $\tilde{\Phi}^a$ is used, then the energy density can be rewritten as
\be
t_{00}=\frac{1}{2}\left[{\l{m^2+\lambda\phi^2}\r\l{(\tilde{\Phi}^0)^2+\tilde{\Phi}^i\tilde{\Phi}^i}\r+m^2\phi^2}\right]\quad,
\ee
which is obviously positive for $\lambda>0$. Note that the terms $m^2\phi^2$ and $m^2\tilde{\Phi}_a\tilde{\Phi}^a$ in the Lagrangian density imply that the dimension of $\phi$ and $\tilde{\Phi}^a$, in terms of the dimension of mass, is $1$. Hence, the coupling constant $\lambda$ is dimensionless and the condition for renormalizability (no inverse mass dimensions of coupling constants) is satisfied. We have thus defined a renormalizable theory with positive energy for a spin zero field with the potential that is second order in field powers. This theory is not equivalent to the $\lambda\phi^4$ theory for a scalar field $\phi$. To see this, note that the Euler--Lagrange field equations following from (\ref{PVLpot}) 
\be
\ba
&(m^2+\lambda \phi^2)\tilde{\Phi}_a=sm\partial_a\phi\quad,\\
&sm\partial_a\tilde{\Phi}^a+m^2\phi=-\lambda\tilde{\Phi}_a\tilde{\Phi}^a\phi
\ea
\ee
can be equivalently expressed as
\be
\ba
&\tilde{\Phi}_a=\frac{sm}{m^2+\lambda\phi^2}\partial_a\phi\quad,\\
&\l{\square+m^2}\r\phi+\lambda\phi^3-\frac{\phi\partial_a\phi\partial^a\phi}{m^2+\lambda\phi^2}=0\quad.
\ea
\ee
Hence, the field $\tilde{\Phi}^a$ is totally determined by $\phi$, and the independent field $\phi$ obyes a nonlinear equation which is clearily different from the one of $\lambda\phi^4$ theory.

The $\lambda\phi^4$ theory could also be introduced in the formalism of a Poincar\'e--vector field by choosing the potential term
$\lambda
{\tilde{\psi}}^T
\l
\begin{array}{cc}
0 & 0 \\
0 & 1
\end{array}
\r
\tilde{\psi}
{\tilde{\psi}}^T
\l
\begin{array}{cc}
0 & 0 \\
0 & 1
\end{array}
\r
\tilde{\psi}
=\lambda\phi^4
$
. However, unlike $\stackrel{\rho}{\mathcal{P}}$ or $\mathbb{P}^a$, the matrix 
$
\l
\begin{array}{cc}
0 & 0 \\
0 & 1
\end{array}
\r
$
does not appear in the theory in a natural way. Therefore, if we decided to use indecomposable faithful representations of the Poincar\'e group to describe physical fields, the more complicated potential constructed in terms of $\mathbb{P}^a$ ought to be given the priority in physical applications. 
Let us now turn to more realistic interactions.

\subsection{The U(1) gauge theory -- electromagnetism}\label{U1}

The theory of a free Poincar\'e--spinor field appears to be just another description of standard fermionic theory. Also, the free theory of a Poincar\'e--vector field of spin $0$ reduces to the description of a usual scalar field in first order formalism, although the criteria of what kinds of interactions seem to be more natural then others depend on the choice of formalism. The hitherto results suggest that perhaps all the physically important fields can be viewed as caring faithful indecomposable representations of the Poincar\'e group (this should be contrasted with the Lorentz group approach in which a non-faithful representation is necessary to describe a scalar field) . Let us now inspect how the minimal coupling procedure for Yang--Mills gauge theories works in both formalisms. In this first part of the paper, we shall consider the simplest example of electromagnetic interaction.
\vskip 0.1 in

\noindent {\bf Poincar\'e--spinor field of spin $1/2$}

The Lagrangian density (\ref{modL0}) is clearly invariant under the global $U(1)$ transformations $\psi\rightarrow e^{i\lambda}\psi$, where $\lambda\in\mathbb{R}$ is a parameter of a transformation. The corresponding conserved Noether current is 
\be
j^a(x)=\ov{\psi}(x)\tilde{\gamma^a}(x)\psi(x)\quad.
\ee
The explicit dependence on $x$ that is involved in $\tilde{\gamma^a}$ may seem to suggest that strange physical effects may occur when the current, possibly coupled to electromagnetic field, is observed by inertial observers that are spatially or temporally translated with respect to each other (not necessarily in relative motion). However, these effects will not occur, since the current reduces to the standard one when expressed in terms of annihilation and creation operators, or just in terms of the Lorentz--transforming field $\tilde{\psi}(x)=\rho^{-1}(x)\psi(x)$.

The local action of $U(1)$ is obtained by allowing $\lambda$ to be a function on space--time. Imposing the invariance of the Lagrangian (\ref{modL0}) under such local action by the replacement
\be
D_a\psi=\partial_a\psi+ieA_a\psi\quad,
\ee
where $e$ is electric charge, 
and adding the pure gauge field part $-\frac{1}{4}F_{ab}F^{ab}$, $F_{ab}:=\partial_a A_b-\partial_b A_a$ one obtains
\be
\mathcal{L}_{em}=\ov{\psi}\left[{\tilde{\gamma^a}(i\partial_a+\mathbb{P}_a)-m}\right]\psi-eA_aj^a-\frac{1}{4}F_{ab}F^{ab}\quad,
\ee
where the electromagnetic one--form transforms under gauge transformations as $A_a\rightarrow A_a-\frac{1}{e}\partial_a\lambda$. Variation with respect to $A_a$ and $\psi$ yields Maxwell's equations and the covariant Dirac equation
\be
\partial_aF^{ab}=ej^b\quad,\quad \left[{\tilde{\gamma^a}(x)(iD_a+\mathbb{P}_a)-m}\right]\psi(x)=\rho(x)\l{i\gamma^aD_a-m}\r\tilde{\psi}(x)=0\quad.
\ee
Hence, the equations, when written in terms of $\tilde{\psi}$, reduce to those of standard electrodynamics of the Dirac field (use invertibility of $\rho(x)$).

\vskip 0.1 in
 
\noindent {\bf Poincar\'e--vector field of spin $0$}

The theory can be reduced to that of a Lorentz transforming field $\tilde{\psi}$ in a completely analogous way to the case of the Dirac field (in both cases this possibility follows from general considerations of Subsection \ref{intgen}). Hence, we shall begin from the start with the Lagrangian density (\ref{PVch}) that is invariant under global $U(1)$ transformations $\tilde{\psi}\rightarrow e^{i\lambda}\tilde{\psi}$. 
The corresponding Noether current is 
\be\label{PVj}
j^a=\frac{im}{\alpha}\l{e^{i\beta}{\tilde{\psi}}^{\dag}\stackrel{\rho}{\mathcal{P}}\mathbb{P}^a\tilde{\psi}-e^{-i\beta}{\tilde{\psi}}^{\dag}\mathbb{P}^{a\dag}\stackrel{\rho}{\mathcal{P}}\tilde{\psi}}\r
\ee
and the $U(1)$ gauge invariant Lagrangian is
\be\label{PVel}
\mathcal{L}_{em}=-\frac{m}{\alpha}\l{e^{i\beta}{\tilde{\psi}}^{\dag}\stackrel{\rho}{\mathcal{P}}\mathbb{P}^a\partial_a\tilde{\psi}+e^{-i\beta}\partial_a{\tilde{\psi}}^{\dag}{\mathbb{P}^a}^{\dag}\stackrel{\rho}{\mathcal{P}}\tilde{\psi}}\r-m^2{\tilde{\psi}}^{\dag}\stackrel{\rho}{\mathcal{P}}\tilde{\psi}-eA_aj^a-\frac{1}{4}F_{ab}F^{ab}\quad.
\ee
Note that in both cases turning on electromagnetism leads to the passage $\mathcal{L}\rightarrow\mathcal{L}-eA_aj^a-\frac{1}{4}F_{ab}F^{ab}$, where $\mathcal{L}$ does not depend on the electromagnetic field. This leads straightforwardly to the interpretation of $ej^a$ as electric current four--vector, as it appears as a source in Maxwell's equations $\partial_aF^{ab}=ej^b$. If standard formalism for a scalar field described by (\ref{Lst}) is used instead, the procedure is slightly awkward. The initial conserved current corresponding to global $U(1)$ symmetry is
\be
J^a_{st}=i\l{\phi^*\partial^a\phi-\phi\partial^a\phi^*}\r
\ee
and the localization $\partial_a\phi\rightarrow D_a\phi=\partial_a\phi+ieA_a\phi$ leads to
\be\label{PVstem}
\mathcal{L}_{st,em}=\partial_a\phi^*\partial^a\phi-m^2\phi^*\phi-eA_aj^a_{st}+e^2A_aA^a\phi^*\phi-\frac{1}{4}F_{ab}F^{ab}\quad,
\ee
where the additional term $e^2A_aA^a\phi^*\phi$ is difficult to interpret at first. However, after the symmetry has been localized, the original current $J^a_{st}$ is no longer conserved. The relevant $U(1)$ Noether current is now
\be
\tilde{j^a}_{st}=i\l{\phi^*\partial^a\phi-\phi\partial^a\phi^*}\r-2eA^a\phi^*\phi\quad.
\ee
The interpretation of $e\tilde{j^a}$ as an electric current is reinforced by varying with respect $A_a$ and deriving Maxwell's equations in the form
\be
\partial_aF^{ab}=e\tilde{j^b}\quad.
\ee
Note that the current $\tilde{j^a}$ explicitly depends on $A_a$ and that the Lagrangian $\mathcal{L}_{st,em}$ is not of the form $\mathcal{L}_{st}-eA_a\tilde{j^a}-\frac{1}{4}F_{ab}F^{ab}$. By the comparison, one should appreciate the elegance and simplicity of the description in terms of a non--standard formalism for the scalar field that is inspired by considering faithful representations of the Poincar\'e group. It should also be noted that the non--standard Lagrangian (\ref{PVel}) does not contain a derivative interaction terms that are present in (\ref{PVstem}) and make the quantization more difficult. Nevertheless, in spite of these aesthetic differences, the resulting theories are classically equivalent: varying (\ref{PVel}) with respect to ${\tilde{\Phi}}^{a*}$ (first four entries of ${\tilde{\psi}}^{\dag}$) leads to ${\tilde{\Phi}}_a=\frac{i}{m}e^{i\beta}D_a\phi$ which, when inserted back to $\mathcal{L}_{em}$, reduces it to $\mathcal{L}_{st,em}$. Note that the relation between ${\tilde{\Phi}}$ and $\phi$ is $U(1)$--covariant and involves the electromagnetic field. Hence, also the current (\ref{PVj}) acquires the $A_a$-dependence when field equations are partially solved (in fact, it is then equal to $\tilde{j^a}_{st}$).

\section{Conclusions}\label{conc}

It is possible to construct a theory of quantum fields that transform according to faithful representations of the Poincar\'e group (we shall refer to such fields as Poincar\'e fields, whereas the fields that do not feel translations will be referred to as Lorentz fields).

For every Poincar\'e field $\psi(x)$, transforming under the representation $\rho(\Lambda,b)$, there exists a corresponding Lorentz field $\tilde{\psi}(x)=\rho^{-1}({\bf 1},x)\psi(x)$. It was shown that the theory of a free field $\psi$ is equivalent to that of $\tilde{\psi}$. This is true for both classical and quantum theory, since the field $\rho({\bf 1},x)$ providing the difference is purely classical in nature, i.e. it does not hide any operators acting on the Fock space. 

This observation does not mean that considering the Poincar\'e fields cannot result in new physical predictions, since there exist indecomposable representations of the Poincar\'e group $\rho(\Lambda,b)$ that include decomposable representations $\rho(\Lambda,0)$ of the Lorentz group. One would then not expect $\rho(\Lambda,0)$ to describe an elementary system, within the standard framework for QFT (unless the field components corresponding to different constituent irreducible representations of $\rho(\Lambda,0)$ are mixed by discrete symmetries). Rather, one would consider the constituent irreducible representations and assign the unique, and in general distinct, values of spin to them. If, on the other hand, the faithful representations of the Poincar\'e group are treated seriously, then $\rho(\Lambda,b)$, being indecomposable, should be considered as describing an elementary system. The conditions of consistency between $\rho(\Lambda,b)$ and $U(\Lambda,b)$ should be imposed, where $U(\Lambda,b)$ is an irreducible unitary representation of the Poincar\'e group on the Fock space, corresponding to the unique values of mass $m$ and spin $j$. In this way, a description of particles characterized by $m$ and $j$ is achieved, which is alternative to the standard description. Such a situation was depicted on the example of the Poincar\'e--vector field, which provides the non--standard description of spin-less particles.

It seems that all the physically important fields can be described in terms of faithful indecomposable representations of the Poincar\'e group, i.e. the non--faithful representations are not needed, if the Poincar\'e group is used. Alternatively, if the Lorentz group is employed, the non--faithful (trivial) representation is necessary to account for the presence of spin-less particles in nature. Although the $j=0$ condition can be satisfied for the faithful Lorentz vector representation, all the components of the resulting field are then derivatives of a scalar field, which makes extremely difficult (if possible) providing a satisfactory Lagrangian formulation. Such a theory of spin-less particles is therefore necessarily incomplete. This difficulty is not present in the Poincar\'e--vector description, a Lagrangian formulation of which was presented in Subsection \ref{Poinv}.   

As could be expected, the important features of QFT, such as the relations between spin and statistics, remain unaffected by the generalization that is considered. In particular, the Poincar\'e--spinors were shown to be fermions, whereas the Poincar\'e--vector field of spin zero was shown to describe bosons.

The question of an equivalence between the new and the standard way of describing particles becomes a subtle issue when the interactions are introduced. Certainly, the theory of a field $\psi$ is still equivalent to that of $\tilde{\psi}$, since the difference provided by $\rho(x)$ is compensated by the necessarily new way of constructing the coefficients $g_{l_1\cdots l_N}$ that are involved in the general method of obtaining interaction terms (see Subsection \ref{intgen}). Therefore, the theory of the interacting Poincar\'e--spinor field is equivalent to the standard theory of the spinor field. For the Poincar\'e--vector field of spin zero, where the associated Lorentz field $\tilde{\psi}$ provides a non--standard description of spin-less particles, the situation is more interesting. It was argued in Subsection \ref{intgen} that the most natural potential term for $\tilde{\psi}$ that is of fourth order in field powers yields a theory with positive energy that satisfies a necessary condition for renormalisability, which is not equivalent to the ``phi to the fourth'' theory for standard scalar field. This simple example shows that the potentials that arise naturally within one formalism may seem artificial within the second one, and vice versa. It would be interesting to inspect more elaborate indecomposable representations of the Poincar\'e group and the potential terms that naturally emerge from them.

In the case of the two examples that were considered, minimal coupling of electromagnetism leads to the theories that are classically equivalent for both formalisms (see Subsection \ref{U1}), although the identification of the physical electric current carried by particles seems to be more obvious in the Poincar\'e formalism. The analysis and the conclusions can be readily generalized to the case of nonabelian gauge theories whose gauge group actions are independent of the action of the Poincar\'e group. The case of gravity, for which the gauge group is the Poincar\'e group itself, is going to be considered separately in the forthcoming article.

\section{Possible directions for further investigations}\label{further}

The fields corresponding to other indecomposable representations of the Poincar\'e algebra should be considered (see Appendix \ref{A3}). It would also be interesting to complete the quantization of the theory defined by Lagrangian density (\ref{PVLpot}) and find out whether there are differences in physical predictions, when compared to those of the standard ``phi to the fourth'' theory for scalar field.  

It is tempting to postulate that all the possible solutions to the Weinberg consistency conditions that exist for a given indecomposable representation of the Poincar\'e group describe particles which are somehow ``related'', although the precise meaning of this relation cannot be given at this stage of research. For example, the massive scalar field would then be related to the massive vector field. When looking at numbers designating the dimensions of all the representations of $so(3)$ that are included in a given indecomposable representation of the Poincar\'e algebra $iso(1,3)$ (see \cite{LG}), it is clear that fermions cannot be related to bosons in this way by means of representations listed in \cite{LG} and hence it is not possible to ``rediscover'' supersymmetry by considering the content of indecomposable representations considered in \cite{LG}. But the authors of \cite{LG} do not claim that the list of indecomposable representations of $iso(1,3)$ that they present is complete. It would be extremely interesting to find an indecomposable representation of $iso(1,3)$ that contains both even and odd dimensional representations of $so(3)$, or to prove that such representations do not exist.

As far as the fundamental interactions are concerned, the possibility of consistent inclusion of gravity interpreted as a gauge theory needs to be considered.

Another important generalization of the investigations of this paper is to account for the case of massless particles properly.

\section*{Acknowledgements}
I wish to thank to P. Chankowski, W. Kaminski, J. Lewandowski and K. Meissner for helpful comments.
This work was partially supported by the Foundation for Polish Science, grant ''Master''.

\section{Appendix: Notation and conventions}\label{A1}

Throughout the paper $a,b,\dots$ are orthonormal tetrad indices and $\mu,\nu,\dots$ correspond to
a holonomic frame. For inertial frame of flat Minkowski space,
which is both holonomic and orthonormal, we use $a,b,\dots\in\{0,1,2,3\}$ for the whole space--time and $i,j,\dots\in\{1,2,3\}$ for the spatial section.
The metric components in an orthonormal tetrad basis $\tilde{e}_a$ are
 $g\l{\tilde{e}_a,\tilde{e}_b}\r=(\eta_{ab})=diag(1,-1,-1,-1)$ and the dual basis of one--form fields (the cotetrad) is denoted by $e^a$ 
(hence, $e^a(\tilde{e}_b)={\delta^a}_b$). Lorentz
 indices are shifted by $\eta_{ab}$. $\epsilon=e^0\wedge e^1\wedge
e^2\wedge e^3$ denotes the canonical
volume four--form  whose components in orthonormal tetrad basis obey $\epsilon_{0123}=-\epsilon^{0123}=1$.
The Hodge star action on external products of orthonormal cotetrad
one--forms is given by
\beq
\star e_a=\frac{1}{3!}\epsilon_{abcd}e^b\w e^c\w e^d \ , \quad 
\star \l{e_a\w e_b}\r=\frac{1}{2!}\epsilon_{abcd}e^c\w e^d \ , \quad 
\star \l{e_a\w e_b\w e_c}\r=\epsilon_{abcd}e^d \ ,
\eeq
which by linearity determines the action of $\star$ on any differential
form.

\section{Appendix: Noether theorem}\label{A2}

Let  
\be
S[\Phi^A]=\int\mathcal{L}\l{\Phi^A,\partial_{\mu}\Phi^A}\r\d^4 x
\ee
represent  the action of a field theory on a smooth manifold $\mathcal{M}$ (which is not necessarily the Minkowski space). Here $x^{\mu}$ are arbitrary coordinates and hence $\mathcal{L}$ is a scalar density. Let $\mathcal{T}$ be the target
space in which the collection of fields $\Phi$ take its values. 
Consider a Lie group $\mathcal{G}$ that acts on $\mathcal{T}$ as a group of transformations. Let 
\be\label{symtr}
\ba
\Phi^A\longrightarrow \Phi'^A=\Phi^{A}+\delta \Phi^{A} 
\ea
\ee
represent the infinitesimal form of the action of $\mathcal{G}$ on
$\mathcal{T}$. The transformations are called 
{\it symmetry transformations} if they do not change the
action, up to possibly surface terms (and thus leave the form of field
equations invariant). This is equivalent to
\be\label{symcond}
\frac{\partial\mathcal{L}}{\partial\Phi^A}\delta\Phi^A+ 
\frac{\partial \mathcal{L}}{\partial(\partial_{\mu}\Phi^A)}
\partial_{\mu}\delta\Phi^A=\partial_{\mu}W^{\mu}\quad ,
\ee
where $W^{\mu}$ is a vector density. This can be further expressed as
\beq
\partial_{\mu}j^{\mu}=
\l{\partial_{\mu} 
\frac{\partial \mathcal{L}}{\partial(\partial_{\mu}\Phi^A)}-\frac{\partial\mathcal{L}}{\partial\Phi^A}}\r
\delta\Phi^A \quad ,
\eeq
where 
\be\label{j}
j^{\mu}=\frac{\partial\mathcal{L}}{\partial(\partial_{\mu}\Phi^A)}\delta\Phi^A-W^{\mu} \quad 
\ee
is a Noether current associated to the symmetry transformation (\ref{symtr}), which
is clearly conserved, i.e. $\partial_{\mu}j^{\mu}=0$, if the
Euler--Lagrange equations for fields are satisfied.

An interesting class of transformations in Minkowski space is constituted by these transformations that act on both the fields and the Minkowskian coordinates $y^a$. The discussion of Noether theorem presented above applies to this case if the coordinates $y^a(x)$ are interpreted as additional fields on space--time. This way of viewing Minkowskian coordinates appears to be very convenient in PGT. The set of all fields $\Phi^A$ consists then of matter fields $\phi^m$ and the so called Poincar{\'e} coordinates $y^a$. The Lagrangian density is of the form
\be
\mathcal{L}=\pounds \mathrm{det} J\quad,
\ee 
where $\pounds$ is a scalar part of $\mathcal{L}$ (which coincides with $\mathcal{L}$ in Minkowskian coordinates) and $J^a_{\mu}:=\partial_{\mu}y^a$ is the Jacobi matrix. In this case, the conserved current (\ref{j}) can be rewritten in a more convenient form as
\be\label{rmj}
\mathrm{j}^a:=\frac{1}{\mathrm{det} J}J^a_{\mu}j^{\mu}=\frac{\partial\pounds}{\partial(\partial_a\phi^m)}\delta\phi^m-
\left[{\frac{\partial\pounds}{\partial(\partial_a\phi^m)}\partial_b\phi^m-\delta^a_b\pounds}\right]\delta y^b-
\frac{1}{\mathrm{det} J}J^a_{\mu}W^{\mu} \quad .
\ee
In the calculation above the identities 
\be
\frac{\partial\pounds}{\partial(\partial_{\mu}\phi^m)}=J^{\mu}_a \frac{\partial\pounds}{\partial(\partial_a\phi^m)}\quad,\quad 
\frac{\partial\pounds}{\partial(\partial_{\mu}y^a)}=-J^{\mu}_b\partial_a\phi^m \frac{\partial\pounds}{\partial(\partial_b\phi^m)}\quad,\quad 
\frac{\partial\mathrm{det}J}{\partial(\partial_{\mu}y^a)}=\mathrm{det}JJ^{\mu}_a
\ee 
where used. Note that $\mathrm{j}$ is a vector field (not a vector density), whose components in the basis $\partial_{\mu}$ are 
$\mathrm{j}^{\mu}=J^{\mu}_a\mathrm{j}^a=\frac{1}{\mathrm{det} J}j^{\mu}$. Hence, if the Euler--Lagrange equations hold, then $0=\partial_{\mu}\l{\mathrm{det}J\,\,\mathrm{j}^{\mu}}\r=\mathrm{det}J\nabla_{\mu}\mathrm{j}^{\mu}$, where $\nabla$ is the Levi--Civita connection of the Minkowski metric. Therefore, $\nabla_{\mu}\mathrm{j}^{\mu}=\nabla_a\mathrm{j}^a=0$. In Minkowskian coordinates the covariant derivative reduces to partial derivative and hence (\ref{rmj}) is also conserved, i.e. $\partial_a\mathrm{j}^a=0$.

Let us now assume that the Lagrangian density is invariant\footnote{
This means that the transformations are symmetries and the corresponding vector density $W^{\mu}$ is equal to zero.
} under the action of the group of space--time translations that acts on $y^a$ as $y^a\rightarrow y^a+\lambda b^a$, where $\lambda $ is an infinitesimal parameter of a transformation and $b$ is an element of $\mathbb{R}^4$. In conventional field theory, the matter fields do not transform under translations, so $\delta\phi^m=0$ and $\delta y^a=\lambda b^a$ imply that the canonical energy--momentum tensor in the form
\be
{t_b}^a=\frac{\partial\pounds}{\partial(\partial_a\phi^m)}\partial_b\phi^m-\delta^a_b\pounds
\ee
is conserved, $\partial_a {t_b}^a=0$. Let us now consider the modified Dirac field, with the Lagrangian given by (\ref{modL0}). Note however that the letter $x$ that appears there ought to be replaced by $y$ according to the notation we use here, since it refers to Minkowskian coordinates. Note also that $\mathcal{L}_0\equiv\pounds_0$ and that $\psi$ and $\ov{\psi}$ have to be considered as independent fields. Since $\delta\psi=i\lambda b\cdot\mathbb{P}\psi$ and $\delta\ov{\psi}=-i\lambda \ov{\psi}b\cdot\mathbb{P}$ (think of $\psi$ and $\ov{\psi}$ as a column and raw matrix respectively), it follows that
\be
\mathrm{j}^a=-\lambda b^b\l{\ov{\psi}\tilde{\gamma}^a(y)\mathbb{P}_b\psi+i\ov{\psi}\tilde{\gamma}^a(y)\partial_b\psi-\delta^a_b\mathcal{L}_0}\r
\ee
and hence the appropriate energy--momentum tensor for the field $\psi$ is
\be
{t_b}^a=\ov{\psi}\tilde{\gamma}^a(y)\mathbb{P}_b\psi+i\ov{\psi}\tilde{\gamma}^a(y)\partial_b\psi-\delta^a_b\mathcal{L}_0\quad.
\ee
The presence of the first component makes this expression look differently from the conventional energy--momentum tensor for the Dirac field. Recall, however, that $\psi=\rho(y)\tilde{\psi}$, where $\rho(y)=\exp\l{iy\cdot\mathbb{P}}\r$ and $\tilde{\psi}$ is the usual Dirac field that transforms trivially under translations. Since 
$i\ov{\psi}\tilde{\gamma}^a(y)\partial_b\psi=i\ov{\tilde{\psi}}\gamma^a\partial_b\tilde{\psi}-\ov{\tilde{\psi}}\gamma^a\mathbb{P}_b\tilde{\psi}$, it follows that ${t_b}^a$ is the standard energy--momentum tensor when expressed in terms of $\tilde{\psi}$. Similarly, the conserved current that is related to the symmetry under the change of a phase of $\psi$ is $\ov{\psi}\tilde{\gamma}^a(y)\psi=\ov{\tilde{\psi}}\gamma^a\tilde{\psi}$. Therefore, these currents will express trough the annihilation and creation operators in exactly the same way as in the standard theory.

Certainly, the interpretation of Minkowskian coordinates as fields, which is useful in PGT, is not necessary to discuss Noether theorem (see e.g. the Appendix of \cite{Kazm1} for more conventional approach).

\section{Appendix: Indecomposable, faithful, finite--dimensional representations of the Poincar\'e group}\label{A3}

Indecomposable representations of the Poincar\'e group, both finite and infinite, have been studied by mathematicians \cite{LG}\cite{Pan}. In \cite{LG}, the infinite--dimensional master representation was constructed on the space of universal enveloping algebra that induces representations on the invariant subalgebra of translations, the ``lowering'' algebra $\Omega_-$ and the ``raising'' algebra $\Omega_+$. Finite--dimensional representations are obtained for these three possibilities on appropriate quotient spaces. They are further subdivided into the cases A1, A2, A3, A4, B1, B2, B3. Without going into details, we shall explain how to use \cite{LG} to obtain quickly explicit matrix form of generators. For definiteness, let us consider the representations on $\Omega_-$, case B3 (other possibilities can be analized similarily). To set up a representation one needs to
\begin{enumerate}
 \item choose $M\in\mathbb{N}_+$ ,
 \item choose $n\in\mathbb{N}_+$ such that $n<M$ ,
 \item chose $q_c\in\mathbb{N}$ such that $q_c\leq M-n$ (this parameter signifies the number of irreducible representations of the Lorentz subalgebra that are contained in the representation of the Poincar\'e algebra) .
\end{enumerate}
The basis of the linear space on which the representation act is then given by the formula (4.2) of \cite{LG}. To calculate the action of the generators on this basis use (3.4)\cite{LG}. The parameters $\Lambda_1$ and $\Lambda_2$ that appear in it are given by $\Lambda_1=M$, $\Lambda_2=\pm in$ (for the B3 case) and $\alpha_{Nq}$, $\beta_{Nq}$, $\delta_{Nq}$, $\gamma_{Nq}$ are defined below (3.4)\cite{LG}. Any time You get from (3.4)\cite{LG} an element $y^m_{Nq}$ with the values of $m$, $N$ or $q$ that do not belong to the range established by (4.2)\cite{LG}, just set this element to zero. In this way, a definite matrix forms of the generators $h_+$, $h_-$, $h_3$, $p_+$, $p_-$, $p_3$, $k_+$, $k_-$, $k_3$, $k_0$ can be obtained. The familiar generators of translations and Lorentz rotations are then given by
\be
\ba
&\mathbb{J}^{01}=\frac{1}{2}\l{p_-+p_+}\r ,\quad
\mathbb{J}^{02}=\frac{1}{2}\l{p_--p_+}\r ,\quad
\mathbb{J}^{03}=p_3 ,\\
&\mathbb{J}^{12}=-h_3 ,\quad
\mathbb{J}^{13}=-\frac{i}{2}\l{h_--h_+}\r ,\quad
\mathbb{J}^{23}=-\frac{1}{2}\l{h_-+h_+}\r ,\\
&\mathbb{P}^0=k_0 ,\quad
\mathbb{P}^1=\frac{i}{2}\l{k_-+k_+}\r ,\quad
\mathbb{P}^2=\frac{1}{2}\l{k_--k_+}\r ,\quad
\mathbb{P}^3=i k_3 .
\ea
\ee
These generators obey the standard commutation relations
\be\label{transcom}
\ba
&[\mathbb{P}^a,\mathbb{P}^b]=0\quad, \\
&[\mathbb{P}^a,\mathbb{J}^{cd}]=-i\l{\eta^{ac}\mathbb{P}^d-\eta^{ad}\mathbb{P}^c}\r\quad, \\ 
&[\mathbb{J}^{ab},\mathbb{J}^{cd}]=
-i\l{\eta^{ad}\mathbb{J}^{bc}+\eta^{bc}\mathbb{J}^{ad}-\eta^{bd}\mathbb{J}^{ac}-\eta^{ac}\mathbb{J}^{bd}}\r\quad .
\ea
\ee

As an example, choose $M=1$, $n=0$, $q_c=1$. The basis consists of the elements $\{y^0_{00},y^1_{00},y^2_{00},y^0_{10},y^0_{11}\}$. The application of (3.4)\cite{LG} then yields the generators 
\be
\ba
&h_3=\l{
\begin{array}{ccccc}
0 & 0 & 0 & 0 & 0 \\
0 & 1 & 0 & 0 & 0 \\
0 & 0 & 0 & 0 & 0 \\
0 & 0 & 0 & -1 & 0 \\
0 & 0 & 0 & 0 & 0 
\end{array}
}\r, \ 
h_+=\l{
\begin{array}{ccccc}
0 & 0 & 0 & 0 & 0 \\
0 & 0 & 0 & 0 & 0 \\
0 & 2 & 0 & 0 & 0 \\
0 & 0 & 2 & 0 & 0 \\
0 & 0 & 0 & 0 & 0 
\end{array}
}\r, \ 
h_-=\l{
\begin{array}{ccccc}
0 & 0 & 0 & 0 & 0 \\
0 & 0 & 1 & 0 & 0 \\
0 & 0 & 0 & 1 & 0 \\
0 & 0 & 0 & 0 & 0 \\
0 & 0 & 0 & 0 & 0 
\end{array}
}\r, \\ 
&p_3=\l{
\begin{array}{ccccc}
0 & 0 & 1 & 0 & 0 \\
0 & 0 & 0 & 0 & 0 \\
-1 & 0 & 0 & 0 & 0 \\
0 & 0 & 0 & 0 & 0 \\
0 & 0 & 0 & 0 & 0 
\end{array}
}\r, \ 
p_+=\l{
\begin{array}{ccccc}
0 & -2 & 0 & 0 & 0 \\
0 & 0 & 0 & 0 & 0 \\
0 & 0 & 0 & 0 & 0 \\
-2 & 0 & 0 & 0 & 0 \\
0 & 0 & 0 & 0 & 0 
\end{array}
}\r, \ 
p_-=\l{
\begin{array}{ccccc}
0 & 0 & 0 & 1 & 0 \\
1 & 0 & 0 & 0 & 0 \\
0 & 0 & 0 & 0 & 0 \\
0 & 0 & 0 & 0 & 0 \\
0 & 0 & 0 & 0 & 0 
\end{array}
}\r, \\ 
&k_-=\l{
\begin{array}{ccccc}
0 & 0 & 0 & 0 & 0 \\
0 & 0 & 0 & 0 & 1 \\
0 & 0 & 0 & 0 & 0 \\
0 & 0 & 0 & 0 & 0 \\
0 & 0 & 0 & 0 & 0 
\end{array}
}\r, \ 
k_0=\l{
\begin{array}{ccccc}
0 & 0 & 0 & 0 & -1 \\
0 & 0 & 0 & 0 & 0 \\
0 & 0 & 0 & 0 & 0 \\
0 & 0 & 0 & 0 & 0 \\
0 & 0 & 0 & 0 & 0 
\end{array}
}\r, \ 
k_3=\l{
\begin{array}{ccccc}
0 & 0 & 0 & 0 & 0 \\
0 & 0 & 0 & 0 & 0 \\
0 & 0 & 0 & 0 & -1 \\
0 & 0 & 0 & 0 & 0 \\
0 & 0 & 0 & 0 & 0 
\end{array}
}\r, \\ 
&k_+=\l{
\begin{array}{ccccc}
0 & 0 & 0 & 0 & 0 \\
0 & 0 & 0 & 0 & 0 \\
0 & 0 & 0 & 0 & 0 \\
0 & 0 & 0 & -2 & 0 \\
0 & 0 & 0 & 0 & 0 
\end{array}
}\r. 
\ea
\ee
This representation is equivalent to the Poincar\'e--vector representation discussed in this paper\footnote{
Note that by $h_3, h_+,\dots$ we mean the matrices representing generators in a particular representation. These are denoted by $\rho(h_3), \rho(h_+),\dots$ in ref.\cite{LG}. But in this article $\rho$ is reserved for the representation of the universal covering of the Poincar\'e group and therefore we are not using it here.
}. To see this, introduce the invertible matrix 
\be
X=
\l{
\begin{array}{ccccc}
2 i\alpha & 0 & 0 & 0 & 0 \\
0 & -2 \alpha & 0 & \alpha & 0 \\
0 & -2i\alpha & 0 & -i\alpha & 0 \\
0 & 0 & 2 \alpha & 0 & 0 \\
0 & 0 & 0 & 0 & 2 
\end{array}
}\r
\ee 
and verify that the relations
\be
\ba
&X^{-1}\mathbb{J}^{01}X= \frac{1}{2}\l{p_-+p_+}\r  ,\quad
X^{-1}\mathbb{J}^{02}X=\frac{1}{2}\l{p_--p_+}\r ,\quad
X^{-1}\mathbb{J}^{03}X=p_3 ,\\
&X^{-1}\mathbb{J}^{12}X=-h_3 ,\quad
X^{-1}\mathbb{J}^{13}X=-\frac{i}{2}\l{h_--h_+}\r ,\quad
X^{-1}\mathbb{J}^{23}X=-\frac{1}{2}\l{h_-+h_+}\r ,\\
&X^{-1}\mathbb{P}^0X=k_0 ,\quad
X^{-1}\mathbb{P}^1X=\frac{i}{2}\l{k_-+k_+}\r ,\quad
X^{-1}\mathbb{P}^2X=\frac{1}{2}\l{k_--k_+}\r ,\quad
X^{-1}\mathbb{P}^3X=i k_3 
\ea
\ee
are satisfied, where $\mathbb{P}^a$ and $\mathbb{J}^{ab}$ are the generators of the Poincar\'e--vector representation (\ref{PVrepr}).

\end{document}